\def\revtex{1}

\ifx\revtex\undefined

\documentclass[entropy,article,submit,oneauthor,pdftex]{Definitions/mdpi}

\firstpage{1}
\pubvolume{1}
\issuenum{1}
\articlenumber{0}
\pubyear{2021}
\copyrightyear{2021}
\datereceived{}
\dateaccepted{}
\datepublished{}
\hreflink{https://doi.org/} 

\Title{Functional Epistemology ``Nullifies'' Dyson's Rebuttal of Perturbation Theory}

\TitleCitation{Open access multistatic radar measurements}

\Author{Karl Svozil $^{1}$\orcidA{}}

\AuthorNames{Karl Svozil}

\AuthorCitation{Svozil, K.}

\address[1]{%
$^{1}$ \quad Institute for Theoretical Physics, TU Wien, Wiedner Hauptstrasse 8-10/136, 1040 Vienna,  Austria; svozil@tuwien.ac.at; \url{http://tph.tuwien.ac.at/~svozil}}

\corres{Correspondence: svozil@tuwien.ac.at}

\abstract{Functional epistemology is about ways to access functional objects by using varieties of methods and procedures. Not all such means are equally capable of reproducing these functions in the desired consistency and resolution. Dyson's argument against the perturbative expansion of quantum field theoretic terms, in a radical form (never pursued by Dyson), is an example of epistemology taken as ontology.}


\DeclareFontFamily{U}{bbold}{}
\DeclareFontShape{U}{bbold}{m}{n}
 {
  <-5.5> s*[1.069] bbold5
  <5.5-6.5> s*[1.069] bbold6
  <6.5-7.5> s*[1.069] bbold7
  <7.5-8.5> s*[1.069] bbold8
  <8.5-9.5> s*[1.069] bbold9
  <9.5-11> s*[1.069] bbold10
  <11-15> s*[1.069] bbold12
  <15-> s*[1.069] bbold17
 }{}

\begin{document}

\else
\documentclass[%
      reprint,
   twocolumn,
 amsmath,amssymb,
 aps,
 pra,
  longbibliography,
 ]{revtex4-2}

\usepackage[dvipsnames]{xcolor}

\usepackage{amssymb,amsthm,amsmath,bm}

\usepackage{tikz}
\usetikzlibrary{calc,decorations.pathreplacing,decorations.markings,positioning,shapes,snakes}

\usepackage[breaklinks=true,colorlinks=true,anchorcolor=blue,citecolor=blue,filecolor=blue,menucolor=blue,pagecolor=blue,urlcolor=blue,linkcolor=blue]{hyperref}
\usepackage{url}

\ifxetex
%
%
\usepackage{fontspec}
\usepackage{fontspec}
\setmainfont{Garamond}
\setsansfont{Garamond}
\fi

\usepackage{mathbbol} 

\begin{document}

\title{Functional Epistemology ``Nullifies'' Dyson's Rebuttal of Perturbation Theory}

\author{Karl Svozil}
\email{svozil@tuwien.ac.at}
\homepage{http://tph.tuwien.ac.at/~svozil}

\affiliation{Institute for Theoretical Physics,
TU Wien,
Wiedner Hauptstrasse 8-10/136,
1040 Vienna,  Austria}

\date{\today}

\begin{abstract}
Functional epistemology is about ways to access functional objects by using varieties of methods and procedures. Not all such means are equally capable of reproducing these functions in the desired consistency and resolution. Dyson's argument against the perturbative expansion of quantum field theoretic terms, in a radical form (never pursued by Dyson), is an example of epistemology taken as ontology.
\end{abstract}


\maketitle

\fi

\section{Functional epistemology}

Notwithstanding metaphysical preferences about ontological realism, also known as Platonism---asserting that
{\it ``some  [[mathematical or physical objects or]] entities sometimes exist
without being experienced by any finite mind''}~\cite{stace,Parsons1995}---versus mathematical
nominalism---claiming that mathematical entities such as numbers and functions do not exist,
quasi {\it ``a subject with no object''}~\cite{Burgess1999}---every
application of mathematical formalism requires some operational access to these objects and entities.
In a broader perspective this can also be seen as semantics in need for a syntactical formalization.

One important aspect of access is a representation of functional objects and entities
that in some formal form correspond to important aspects of those objects and entities.
Nevertheless, although representations vary---spanning a wide range of efficacies and deficiencies---they
should not be confused with the respective mathematical objects or entities.

The original informal conception of function $y = f(x)$ was that of a unique association of an output ``value'' $y$ given an input ``argument'' $x$.
More recent conceptions consider ordered pairs $\left( x , y \right)$, where again $x$ stands for argument(s) and $y$ for unique value(s);
in particular, there must not be two pairs $\left( x , y \right)$ and $\left( x , y' \right)$ with $y \neq y'$.

This naive functional conception was challenged by G\"odel's, Kleenee's, and Touring's formalization
of what functional ``access'' means; for instance, in the form of paper and pencil operations
on a ``paper machine''~\cite{Turing-Intelligent_Machinery}.
These developments closely followed
the paradigm change from Cantor's naive set theory~\cite{Halmos1974-naiveset}
to axiomatic set theories~\cite{HrbacekJech1999};
for instance, Zermelo-Fraenkel set theory.
Indeed, from a foundational perspective, it might not get worse: Rice's theorem, usually proven by reduction to the halting problem,
states that any nontrivial semantic property of a computable function (evaluated by a Turing machine) is undecidable.

Therefore, due to incompleteness and related theorems,
for the sake of formalization, these earlier intuitive and heuristic perceptions of functional performance
had to be modified and restricted.
The current formalization of functions is in terms of ``desirable'' properties, such as,
in particular, effective computability.
This has resulted in the abandonment of functional totality---the pretension that any (every) arbitrary argument $x$ can be associated with a (unique) value $y$---in favor of partiality:
certain functions, such as predictors of the large-scale performance of deterministic systems, need not have
a value accessible by some algorithm---in short,
there is a difference between determinism and predictability~\cite{suppes-1993}.
In such a regime, it can no longer be maintained that the value $y$ exists ---that is,
can be obtained or accessed by some algorithm or computation.

Typical examples of such ``critical'' functions are Turing's halting function,
or Specker's theorems of recursive analysis~\cite{specker57,specker-ges,kreisel},
or Chaitin's ``halting probability'' $\Omega$ in terms of its bitwise expansion~\cite{calude-dinneen06}.
Physical realizations have been, for instance, by reduction to the halting problem~\cite{Yanofsky2016}, suggested in terms of
undecidable classical dynamics~\cite{moore,Bennett1990ud}, $N$-body problems~\cite{svozil-2007-cestial},
or spectral gaps~\cite{cubit-15}.

It may happen that a program implemented on a computer that ``is supposed to compute a limit''---and, with finite resources,
even ``accesses a few approximations or bounds thereof''---and yet this limit is uncomputable and algorithmically inaccessible:
Because some resources, such as computing time or space, that are necessary to compute this limit with, say,
precision up to its $n$th bit, grow faster asymptotically than any computable function of $n$~\cite{rado}.
In intuitive algorithmic terms the difference between a total versus a partial function may be
imagined as the distinction between a {\tt DO}--loop (with fixed finite beginning, ending and increments) and a {\tt WHILE}-loop.
The latter {\tt WHILE}-loop may or may not ``take forever'', depending on the respective termination condition.

Another area of partial value assignments is quantum mechanics.
Extensions of the Kochen-Specker theorem
suggest that, relative to the assumption of noncontextuality, only a ``star-shaped''~\cite[Fig.~5]{PhysRevA.89.032109}
(in terms of hypergraphs~\cite{Bretto-MR3077516}
representing individual contexts by smooth curves~\cite{greechie:71})
context can have definite value assignments.
Observables in all other contexts must be value indefinite~\cite{pitowsky:218,hru-pit-2003,2015-AnalyticKS}.

Still another issue of functional epistemology is the means relativity of functional representation.
The same function can have very different representations and encodings; some exhibiting more or less problematic issues.
For the sake of an example, we shall later represent one and the same function in five different forms.

The selection of particular means is often not a matter of choice but one of pragmatism or even desperation.
Especially theoretical physicists are often criticized for their ``relaxed'' stance on formal rigor.
Dirac's introduction of the needle-shaped delta function is often quoted as an example.
Heaviside, in another instance, responded to criticism for his use of the ``highly nonsmooth''
unit step function\cite[p.~9, \S~225]{heaviside-EMT}:
{\em ``But then the
rigorous logic of the matter is not plain! Well, what of that?
Shall I refuse my dinner because I do not fully understand the
process of digestion? No, not if I am satisfied with the result.''}

This, in a nutshell, seems to be the attitude of field theorists
regarding the use of perturbation series:
It is well documented~\cite{PhysRev.85.631,LeGuillou-Zinn-Justin,Vainshtein1964-2002}
that the commonly used power series expansion
which can be rewritten as inverse power series expansion
\begin{equation}
\begin{split}
f( \alpha ) = \alpha a_0+a_1\alpha + a_2\alpha^2+ \cdots \\
= \sum_{n=0}^\infty a_n \alpha^n
= \frac{1}{\alpha}  \sum_{n=0}^\infty \frac{a_n}{\left(\frac{1}{\alpha}\right)^{n+1}}
.
\end{split}
\label{2022-nul-Dyson}
\end{equation}
in terms of the fine structure constant $\alpha$---that is, the square of the) coupling constant---is divergent.

For this power series to converge, there has to be a finite radius of convergence
centered at the origin at (fictitious) value $\alpha=0$,
thereby including (fictitious) nonvanishing negative values $\alpha < 0$
within which $F$ has to be analytic.
However, because if the (fictitious) coupling between like charges becomes negative, and because by tunneling this cannot be ``contained'',
the vacuum becomes unstable due to pair creation, and (fictitiously) disintegrates explosively.
Hence, Dyson concludes,
the power series $f(-\alpha)$ cannot converge and thus cannot be analytic---a complete contradiction to the assumption.

An immediate reaction would be to perceive these coincidences as ``bordering on the mysterious''~\cite{wigner}.
This spirit is corroborated~\cite{landau1906uber} by statements like Carrier's Rule, pointing out that
``divergent series converge faster than convergent series because they don't have to converge.''
However, as already surmised by Dyson, quantitative considerations from partial summationss show~\cite[p.~4]{Bleistein-Handelsman}
that convergent series ``initially''---that is, with only ``a few orders'' added---may ``largely deviate'' from
the true value of the function it encodes (eg, consider the straightforward Taylor expansion of $\sin e^{e^{e^{e^e}}}$), whereas some
associated asymptotic divergent series ``initially'' converges toward
this value: a ``reasonable'' approximation can be obtained by taking
relatively few terms of this divergent series; whereas ``many more'' terms of the
convergent series are needed to achieve that same degree of accuracy.

Current experiences in quantum field theory corroborate this view: although the asymptotic perturbation series have a zero radius of convergence,
it is effectively possible to obtain good agreement between theoretical calculations based on asymptotic series
and experimental results.
As it turns out the terms in these asymptotic series become increasingly accurate as the series is extended,
and hence the error in the truncated series decreases as more terms are included.

This is true, in particular, for the QED contribution to the electron anomalous magnetic moment $g-2$
up to the tenth order~\cite{kinoshita-PhysRevLett.109.111807}, as compared with the experimental
value~\cite{Hanneke-PhysRevLett.100.120801,Hanneke-PhysRevA.83.052122}.
The same applies for the muon anomalous magnetic moment~\cite{kinoshita-PhysRevLett.109.111808,Keshavarzi_2022}.
Likewise, the theoretical predictions~\cite{Janka_2022} of the Lamb shift show similar good agreement with experiments~\cite{Bezginov_2019,Ohayon-PhysRevLett.128.011802}.

However, it is important to keep in mind that asymptotic series eventually diverge as more orders are taken into account.
One way of coping with the apparent asymptotic divergence is the resummation of the respective series,
in particular, Borel (re)summations~\cite{Boyd99thedevil,rousseau-2004,Helling-2012,Costin-2009,ZINNJUSTIN20101454,Costin_Dunne_2017},
which are in some instances capable to reconstruct an analytic function
from its asymptotic expansion~\cite{Bruning-1996}.

\section{Euler's series of 1760 and its multiple representations}

For the sake of an example that exhibits a wide spread of varied (asymptotic) behaviors ``catching''
the same ``ontologic'' function (or, from a nominal point of view, the same ``subject without object''), consider a series
\begin{equation}
s(x)  = x - x^2+2x^3-6x^4 + \ldots
\label{2011-m-ch-dseess}
\end{equation}
mentioned by Euler in a 1760 publication~\cite[\S~6, p.~220]{Euler60}.
As already observed by Euler this series can, in a nominal way, be considered a ``solution'' of
\begin{equation}
\begin{split}
\left(\frac{d}{dx} +\frac{1}{x^2}\right) s(x) = \frac{1}{x};
\end{split}
\label{2011-m-ch-dsee}
\end{equation}
associated with the differential operator $\mathfrak{L}_x = \frac{d}{dx} +\frac{1}{x^2}$.
This first-order ordinary differential equation has an irregular (essential) singularity at $x=0$ because
the coefficient of the zeroth derivative  ${1}/{x^2}$ has a pole of order $2$.
Therefore,~(\ref{2011-m-ch-dsee}) is not of the Fuchsian type, and cannot be subjected to the
Frobenius method of creating convergent power series solutions.

Nevertheless, $s(x)$ can be represented in at least five ways,
differing substantially with respect to convergence and utility for (physical) computation and prediction.
In what follows, these cases will be enumerated: $s(x)$ can be represented by
\begin{itemize}
\item[(i)] a convergent Maclaurin series (Ramanujan found a series which converges even more rapidly) solution~(\ref{2022-nul-ef1}) based on the Stieltjes function;
\item[(ii)] a proper (Borel) summation of Euler's divergent series~(\ref{2022-nul-ef2})~\cite[Equation~(3.3)]{rousseau-2004};
\item[(iii)] quadrature, that is, by direct integration of~(\ref{2022-nul-ef3})~\cite[Equation~(3.3)]{rousseau-2004};
\item[(iv)] evaluating Euler's (asymptotic) divergent series~(\ref{2022-nul-ef4}) to ``optimal order''~\cite[Equations~(2.12)-(2.14)]{rousseau-2004}; and
\item[(v)] evaluating the respective inverse factorial series~(\ref{2022-nul-ef5})~\cite[Equation~(5.7)]{Weniger2010}.
\end{itemize}

Let
%
%
$
S(x)
= \int_0^\infty    {e^{-t}}/{(1+tx)} dt
$
stand for the Stieltjes function
(cf.~\cite[formula~5.1.28, page~230]{abramowitz:1964:hmf} but with $x \mapsto \frac{1}{x}$),
%
%
$
\gamma
=\lim_{n\rightarrow \infty}\left(
\sum_{j=1}^n \frac{1}{j}- \log n
\right) \approx 0.5772
$,
be the Euler-Mascheroni constant~\cite{Sloane_oeis.org/A001620},
%
%
$\Gamma (z,x)$ represent the upper incomplete gamma function~\cite[formula~6.5.3, page~260]{abramowitz:1964:hmf},
%
%
${\cal B}s(x)$ be the Borel transform of $s(x)$,
%
%
$\left(x\right)_{n}=\Gamma \left(x+n\right)/\Gamma \left( x \right) = x \left(x +1\right)\cdots \left( x + j - 1 \right)$
and $(x)_0=1$
be Pochhammer symbols~\cite[Section~24.1.3, page~824]{abramowitz:1964:hmf},  also known as the rising factorial power, and
%
%
${S}_j^{(k)}$
be Sterling numbers of the first kind
that are the polynomial coefficients of the
Pochhammer symbol $(z-j+1)_j$~\cite[Section~24.1.3, page~824]{abramowitz:1964:hmf}; that is,
$ \sum_{k=0}^j {S}_j^{(k)} z^k
= (z-j+1)_j
= (-1)^j  (-z)_j
$
for $j \in \mathbb{N}\cup \{0\}$.
Then
\begin{align}
s(x)
&
=x S(x)
=   e^\frac{1}{x}   \Gamma \left( 0, \frac{1}{x} \right) \nonumber
\\
&
=  - e^\frac{1}{x}   \left[  \gamma - \log x +\sum_{n=1}^\infty \frac{(-1)^n}{n!n x^n} \right]
\label{2022-nul-ef1}
\\
&
=
\int_0^\infty {\cal B}S (y)  e^{-\frac{y}{x}}   dy  \nonumber
\\
&= \qquad \int_0^\infty \frac{ e^{-\frac{y}{x}} }{1+y}    dy= \int_0^\infty \frac{ x e^{-t }}{1+xt}    dt
\label{2022-nul-ef2}
\\
&
=
 \int_0^x  \frac{ e^{ \frac{1}{x} -\frac{1}{t}}}{t} dt
\label{2022-nul-ef3}
\\
&
= \sum_{j=0}^\infty (-1)^j j!  x^{j+1}
\label{2022-nul-ef4}
\\
&
= x \sum_{j=0}^\infty \frac{(-1)^j}{\left(\frac{1}{x}\right)_{j+1}} \sum_{k=0}^j {S}_j^{(k)} k!
.
\label{2022-nul-ef5}
\end{align}

What we can learn from this prototypic example is the wide variety of mathematical representations
associated with one and the same function. Not everybody might agree with all the equality signs
in (\ref{2022-nul-ef1})--(\ref{2022-nul-ef5}) and the legality of the respective methods though,
thereby reflecting a variety of metamathematical stances.

\section{Quantum field theoretical perturbation series need not diverge}

Let us discuss two critical aspects in the derivation of the power series expansion of~(\ref{2022-nul-Dyson}).
One critical step in the derivation of $f$
amounts to interchanging a sum with an integral
in the case of nonuniform convergence of the former~\cite[Sect.~II.A]{PhysRevD.57.1144}.
One may perceive asymptotic divergence as a ``penalty'' for such manipulations.
It may come as a surprise that those calculations performed well for empirical predictions.

Ritt's theorem inspires one strategy to cope with such issues~\cite{Pittnauer-73,Remmert-1991-tocf} by
stating that any (not necessarily asymptotic) divergent power series with arbitrary coefficients can
be converted into nonunique analytic functions. Thereby, every summand is
multiplied with a suitable nonunique {\em ``convergence factor.''}
(Conversely, every analytic function can be approximated by a unique asymptotic series.)

A general regularisation
of divergent series using such convergence factors,
also called {\em cutoff functions}, has been recently introduced by
Tao~\cite[Section~3.7]{Tao-2013}.
The resulting smoothed sums may become uniformly convergent, thereby
allowing interchanging a sum with an integral
and avoiding the aforementioned issues while preserving
inherent properties of the original divergent series.
This is not dissimilar to the use of test functions in the theory of distributions.

A second critical aspect is the expansion of~(\ref{2022-nul-Dyson}) in terms of a power series, and the resulting vanishing of the radius of convergence.
Dyson already mentioned a possible remedy, his
{\em ``Alternative A: There may be discovered a new method
of carrying through the renormalization program, not making use of power series expansions.''}
One such candidate expansion that does not necessarily share the catastrophic fate of the power series caused by the ``explosive disintegration''
of the vacuum state for negative arguments, is an expansion of $f(\alpha )$ in terms of inverse factorial series~\cite{Watson1912,Doetsch1972}
and recently investigated by Weniger~\cite{Weniger2010} as well as O. Costin, R. D. Costin and Dunne~\cite{Costin2016Aug,Costin_Dunne_2017}:
\begin{equation}
f(\alpha)
=
b_0 \frac{0!}{\alpha}
+
b_1 \frac{1!}{\alpha(\alpha +1)}
+
\ldots
= \sum_{n=0}^\infty b_n \frac{n!}{(\alpha)_{n+1}}
,
\label{2022-nul-ifs}
\end{equation}
where again $\left(\alpha\right)_{n+1}$ are Pochhammer symbols.

Stirling numbers of the first kind ${S}_j^{(k)}$ mentioned earlier
serve as ``translations''---that is, as expansions from
an  inverse  power ${1}/{ \alpha ^{n+1}}$ in terms of  inverse  factorial series $( \alpha )_{n+j+1}$:
for $k  \in \mathbb{N} \cup \{0\}$~\cite[Equation~(6), \S~30, p.~78]{Nielsen-Gammafunktion},
\begin{equation}
\begin{split}
\frac{1}{ \alpha ^{n+1}}  = \sum_{j=0}^\infty \frac{(-1)^j}{( \alpha )_{n+j+1}} {S}_{n+j}^{(n)}
.
\end{split}
\label{2022-m-ch-dsngf}
\end{equation}

The respective ``reverse'' expansion of a Pochhammer symbol $( \alpha )_{k+1}$ in terms of  an inverse power series
${1}/{ \alpha ^{n+j+1}}$
for $k  \in \mathbb{N} \cup \{0\}$ and $\vert \alpha \vert >0$~\cite[Equation~(9), \S~26, p.~68]{Nielsen-Gammafunktion}
is given by
\begin{equation}
\begin{split}
\frac{1}{( \alpha )_{n+1}}  = \sum_{j=0}^\infty \frac{(-1)^j}{\alpha^{n+j+1}} {S}_{n+j}^{(n)}
.
\end{split}
\label{2022-m-ch-dsngfinverse}
\end{equation}

Insertion of~(\ref{2022-m-ch-dsngf})  into~(\ref{2022-nul-Dyson}), rearranging the order of the summations through an index shift $m = n+j$
 with $n \ge 0$  and $j \ge 0$, hence  $m \ge 0$ and $j = m - n \ge 0$ and $n \le m$     yields
\begin{equation}
\begin{split}
f\left(  \alpha  \right)
=  \frac{1}{\alpha}  \sum_{n=0}^\infty a_n \sum_{j=0}^\infty \frac{(-1)^j}{\left(\frac{1}{\alpha}\right)_{n+j+1}} \, {S}_{n+j}^{(n)}
\\
=  \frac{1}{\alpha} \sum_{j=0}^\infty \sum_{n=0}^\infty a_n \frac{(-1)^j}{\left(\frac{1}{\alpha}\right)_{n+j+1}} \, {S}_{n+j}^{(n)}
\\
=  \frac{1}{\alpha} \sum_{m=0}^\infty   \frac{(-1)^m}{\left(\frac{1}{\alpha}\right)_{m+1}} \sum_{n=0}^m (-1)^{\pm n} \, {S}_{m}^{(n)} \,  a_n
.
\end{split}
\label{2022-m-ch-dsinv2}
\end{equation}
Therefore, if we define the inverse power series
$
f'(\beta )
= (1/\beta ) f(1/\beta )=\sum_{m=0}^\infty a'_m/\beta^{m+1}
= \sum_{m=0}^\infty b'_m {m!}/(\beta)_{m+1}
$ with $\beta =1/\alpha$,
then, by comparison,
\begin{equation}
\begin{split}
b_m'= \frac{1}{m!} \sum_{n=0}^m (-1)^{m\pm n} \, {S}_{m}^{(n)} \,  a_n'
.
\end{split}
\label{2022-m-ch-comp}
\end{equation}

For the sake of studying the fascinating convergence~\cite{Nielsen-Gammafunktion,landau1906uber,Doetsch1972,Weniger2010} of the inverse factorial series,
note that terms of the form ${n!}/{(z)_{n+1}}$
can, for large $z\rightarrow \infty$,  be estimated  with the help of~\cite[Formula~6.1.47]{abramowitz:1964:hmf}
$\Gamma(z+a)/\Gamma(z+b)= z^{a-b}\left[1 + O\left(\frac{1}{z}\right) \right]$
and for large $n \rightarrow \infty$ as follows:
\begin{equation}
\begin{split}
\frac{n!}{( \alpha )_{n+1}}
= \frac{\Gamma(n+1)}{\left[\Gamma( \alpha +n+1)/\Gamma( \alpha )\right]}
= \frac{\Gamma(n+1)}{\Gamma(n+1+\alpha)} \Gamma( \alpha )  \\
= (n+1)^{-\alpha}\left[1 + O\left(\frac{1}{n+1}\right) \right]( \alpha -1)!
= O \left( n^{-\alpha} \right).
\end{split}
\label{2022-m-ch-estimate}
\end{equation}

Therefore, the inverse factorial series (\ref{2022-nul-ifs})
converges with the possible
exception of the points $\alpha =-m$ with $m \in   \mathbb{N} \cup \{0\}$
(where the Pochhammer symbols in the denominator might vanish)
if and only if the associated Dirichlet series
$\sum_{n=1}^\infty c_n \, n^{-\alpha }$
converges.

Unlike a power series that has a radius of convergence, a Dirichlet series has an abscissa of convergence
$\Re ( \alpha ) > \lambda$, that is, it converges on this half-plane~\cite[\S~58,~255, page~456]{Knoop1996}.
Therefore, the inverse factorial series may converge for all positive coupling constants $\alpha >0$ although
it may diverge for negative values $\alpha < 0$. The physically relevant region lies within the abscissa of convergence.
Even if the inverse power series diverges factorially, the respective inverse factorial series may converge, but this has to be
checked explicitly.

Thus, as already suggested by Dyson~\cite{PhysRev.85.631},
representing quantum field theoretical entities in terms of the
Tomonaga-Schwinger-Feynman-Dyson power series expansion in the coupling constant~\cite{Dysen-49}
may suffice for all practical purposes~\cite{bell-a} so far,
although their divergencies may cause uneasiness for a variety of (pragmatic and formal) reasons.
One should not confuse these field-theoretic entities with their actual representations; that is,
functional ontology with epistemology.
Such considerations might present a positive outlook for an improved theory of convergent perturbation series.
A candidate for such a theory might be inverse factorial expansions exhibiting an abscissa rather than a radius of convergence.

However, a convergence issue encountered in inverse factorial series is the Stokes phenomenon~\cite{Costin2016Aug,Costin_2017}:
the asymptotic behavior of functions need not be uniform in different regions of the complex plane, bounded by (anti-)Stokes lines.
In particular, inverse factorial series may not be suitable for the study of Stokes phenomena if Stokes lines
are present in the right complex half-plane $\Re ( \alpha ) > \lambda $ because of the singularities on these Stokes lines.
One may conjecture that inverse factorials might converge in regions where the associated power series are Borel summable;
yet convergence fails in the presence of Stokes lines.
This would mean that quantum field theories have convergent inverse factorial expansions only in less than four dimensions;
and that this expansion might fail for four dimensions.
Nevertheless, Berry, O Costin, R. D. Costin and Howls have pointed out~\cite[p.~10]{costin2016} that,
although {\it ``typically, classical factorial series have two major limitations: slow
convergence, at best power-like, and a limited domain of convergence (a half
plane which cannot be centered on the asymptotically important Stokes line) $\ldots$
for resurgent functions these limitations can be overcome. Ecall\'e--Borel summable series can be summed by rapidly convergent factorial series.''}

Transseries from Borel-Ecall\'e summations of divergent (power) series offer a method to ``recover''
nonperturbative information from such power series~\cite{Costin_1995,Edgar-2009},
thereby indicating that the divergent perturbative power series expansion contains information of the nonperturbative kind.
The situation is not totally dissimilar from tempered distributions: using test functions with unbounded (noncompact) support
allows the representation and reconstruction of generalized functions by Fourier transforms.

A further method for alternative representations of functions are Pad\'e approximations
by rational functions (of given order) near a specific point.
Pad\'e approximantions offer practical methods of defining
and computating the value of a power series even if such series diverge~\cite{baker_graves-morris_1996}.


\section{Summary}

In this brief expos\'e I have bundled together two ideas:
first, that mathematical objects or entitities such as functions or proofs~\cite{ziegler-aigner} may have very varied representations and realizations.
Not all of them might require comparable means to access them---think of convergence or (asymptotic) divergence, or of partial functions.
Different means might not be equally appropriate or sufficient and necessary for different purposes.

Second, in particular and more specifically, as conjectured already by Dyson,
arguments against the existence or convergence of power expansions of Tomonaga-Schwinger-Feynman-Dyson perturbative quantum field theory
might be ``liftable'' by using other expansion techniques.

For the sake of illustration, suppose for a moment that G\"odel's ``unadulterated'' Platonism~\cite{kreisel-80,Parsons1995} is acceptable.
(An analogous argument can be made within nominalism.)
Then mathematical objects or entities such as functions can be conceptualized by their ontological existence.

However,
on second thought, it is an entirely different, highly nontrivial, issue to ``touch'' or to epistemically access those objects or entities.
We have presented some examples of such access which analytically spread over a wide variety of
(asymptotic) divergent and convergent expressions.

We have, in particular, argued that (asymptotic) divergence
of a particular type of perturbation series based on power series expansions
could be overcome by other methods of perturbative series; in particular,
by inverse factorial series.
This still leaves open the consistent existence of quantum field theory,
but at least it indicates conceivable convergent access to quantum field theoretical objects and functions.

\ifx\revtex\undefined

\funding{This research was funded in whole, or in part, by the Austrian Science Fund (FWF), Project No. I 4579-N. For the purpose of open access, the author has applied a CC BY public copyright licence to any Author Accepted Manuscript version arising from this submission.
}

\acknowledgments{I kindly acknowledge explanations by and considerations with Cristian S. Calude, Alexander Leitsch and Noson S. Yanofsky,
as well as discussions with and suggestions by Thomas Sommer. {\it Mea culpa} if I got them wrong!}

\conflictsofinterest{

The author declares no conflict of interest.
The funders had no role in the design of the study; in the collection, analyses, or interpretation of data; in the writing of the manuscript, or in the decision to publish the~results.}

\else

\begin{acknowledgments}

I kindly acknowledge explanations and considerations with Cristian S. Calude, Alexander Leitsch and Noson S. Yanofsky,
as well as discussions with and suggestions by Thomas Sommer. Their patience with me is highly appreciated. Mea culpa if I got them wrong!

This research was funded in whole, or in part, by the Austrian Science Fund (FWF), Project No. I 4579-N. For the purpose of open access, the author has applied a CC BY public copyright licence to any Author Accepted Manuscript version arising from this submission.

The author declares no conflict of interest.
\end{acknowledgments}

\fi

\ifx\revtex\undefined

\end{paracol}
\reftitle{References}


 \externalbibliography{yes}
 \bibliography{svozil,ufo}

\else


%

\fi
\end{document}